\documentclass[a4paper,fleqn,10pt]{article}
\usepackage{graphicx}
\usepackage{color}

\usepackage[square,numbers,sort&compress]{natbib}
\usepackage{amsmath,amssymb,amsfonts}

\usepackage{graphicx}
\usepackage{array}
\usepackage{threeparttable}
\usepackage{booktabs}
\usepackage{tabularx}
\usepackage{color}
\usepackage{longtable}
\usepackage{chemarrow}
\usepackage{titlesec}
\usepackage{url}
\usepackage{times}
\usepackage{mathptmx}

\usepackage[font={small}]{caption}
\usepackage{amsmath,bm}
\usepackage{anysize}
\marginsize{2cm}{2cm}{0cm}{2cm}

\title{A physical mechanism of heterogeneity in stem cell, cancer and cancer stem cell}

\author{Chong Yu,$^{1,2}$  Qiong Liu,$^{1}$  Cong Chen,$^{4}$ and Jin Wang$^{1,2,3,4*}$\\
	\normalsize{$^1$ State Key Laboratory of Electroanalytical Chemistry } \\
	\normalsize{Changchun Institute of Applied Chemistry, Chinese Academy of Sciences } \\
	\normalsize{Changchun, Jilin 130022, China } \\
	\normalsize{$^2$ University of Science and Technology of China} \\
	\normalsize{$^3$ College of Physics} \\
	\normalsize{Jilin University, Changchun, Jilin 130012, China}\\
	\normalsize{$^4$ Department of Chemistry, Physics \& Applied Mathematics} \\
	\normalsize{State University of New York at Stony Brook } \\
	\normalsize{Stony Brook, NY 11794-3400, USA } \\
	\normalsize{$^{*}$Corresponding Authors: jin.wang.1@stonybrook.edu}\\}
\date{}

\begin{document}
\maketitle



\begin{abstract}
Heterogeneity is ubiquitous in stem cells (SC), cancer cells (CS), and cancer stem cells (CSC). SC and CSC heterogeneity is manifested as diverse sub-populations with self-renewing and unique regeneration capacity. Moreover, the CSC progeny possesses multiple plasticity and cancerous characteristics. Many studies have demonstrated that cancer heterogeneity is one of the greatest obstacle for therapy. This leads to the incomplete anti-cancer therapies and transitory efficacy. Furthermore, numerous micro-metastasis leads to the wide spread of the tumor cells across the body which is the beginning of metastasis. The epigenetic processes (DNA methylation or histone remodification etc.) can provide a source for certain heterogeneity. In this study, we develop a mathematical model to quantify the heterogeneity of SC, CSC and cancer taking both genetic and epigenetic effects into consideration. We uncovered the roles and physical mechanisms of heterogeneity from the three aspects (SC, CSC and cancer). In the adiabatic regime (relatively fast regulatory binding and effective coupling among genes), seven native states (SC, CSC, Cancer, Premalignant, Normal, Lesion and Hyperplasia) emerge. In non-adiabatic regime (relatively slow regulatory binding and effective weak coupling among genes), multiple meta-stable SC, CS, CSC and differentiated states emerged which can explain the origin of heterogeneity. In other words, the slow regulatory binding mimicking the epigenetics can give rise to heterogeneity. Elucidating the origin of heterogeneity and dynamical interrelationship between intra-tumoral cells has clear clinical significance in helping to understand the cellular basis of treatment response, therapeutic resistance, and tumor relapse.

\end{abstract}

\section{Introduction}

Cells are the basis for life. Cells can replicate \cite{Hiratani2008Global} or differentiate \cite{Harmeet2002Isolation}. Cells can switch phenotypes from the stem cells to the differentiated cells. In the process of differentiation and development, mutations and genetic changes are often not significant\cite{Sell2004Stem}. The underlying gene regulatory networks are believed to provide the driving force for differentiation \cite{Li2015CR,Li2018Uncovering,Wenbo2017}. In stem cells,  heterogeneity is often found \cite{Tang2012Understanding,Marusyk2012}. The heterogeneity here obviously does not come from genetic changes such as mutations but must be from other roots. Epigenetics and micro-environments may provide sources for the heterogeneity in stem cells\cite{Wang2015Heterogeneity,Dagogo2017Tumour}.

Cancer is a systemic level disease which involves not only a mixture of tumor cells, but also the microenvironment, signal transduction, extracellular components etc.. In the early studies, genetic mutation was believed to be the main driving force for the cancer initiation and progression\cite{Vogelstein2004Cancer,Kreso2014,Marco2012Intratumor}. Recent studies revealed that cancer is not just a genetic disease, but should be considered as an ecosystem which is under environmental selection\cite{Greaves2012Clonal,Kim2016}. The mutations are more likely to appear when the cell lesions are developed. The mutant cells are subject to epigenetic influences and micro-environmental pressures. They prefer to spread out to other organs. These can lead the mutant cells to acquire different “hallmarks of cancer”\cite{Hanahan2000-57}. As cancer involves epigenetic and micro-environmental influences, the studies on the underlying cancer gene regulatory networks and how the networks change with respect to the epigenetics and environment have received recent attentions \cite{Xu2014Exploring, Li2013PlosComp, Li2014JRSI, Li2015CR}.

CSC is a type of tumor cells which occupy a very small proportion at about 1\%-4\%\cite{BRUCE1963}. Recently, a growing number of evidences revealed that CSC is the main driving force for cancer recurrence, drug resistance and migrate\cite{Wu2011Direct,MaugeriSacca2011}. If the cancerous mutations acquire a specific capacity, the stemness, the progeny may turn out to be the CSCs. If the somatic stem cells acquire disorder characteristics of cancer cells, the progeny maintains the stem cell-like self-renewal capability and possesses the cancerous characteristics. Therefore, the CSC progeny can have various tumor cell phenotypes and the self-renewal ability.

Heterogeneity is one of the most significant contributors to various tumor cell phenotypes. The heterogeneity can be related to epigentics. Without basing the DNA sequence changes, diversity cell phenotypes can still be generated through the epigenetic mechanisms, such as DNA methylation and histone modifications\cite{Baylin2011A}. These will result the gene expression change and the change of the cell phenotypes\cite{Shenker2011}. Epigenetic changes are heritable and can influence gene function directly. The variations can be accumulated. They can then contribute to clonal selection and provide a source for tumor cell heterogeneity.

In this study, we start from a gene regulatory motif involving SC, CS, CSC, and differentiation with two oncogene of cancer (P53 and MDM2), one marker gene of stem cell (OCT4), one marker gene of metastasis (ZEB) and two micro-RNAs (miR-145 and miR-200) which are crucial in the regulations.  We develop a landscape framework to quantify the heterogeneity of SC, CSC and cancer. To explore the epigenetic aspects we vary the time scales of the regulatory process of  protein binding/unbinding with respect to protein synthesis/degradation rate.

In our previous work, we have explored the fast adiabatic binding/unbinding gene regulation regime in details \cite{Chong2018Quantification}. In adiabatic regime, seven steady states are emerged, Normal, Premalignant, Cancer, SC, CSC, Lesion and Hyperplasia state. In this work, we mainly study slow fast non-adiabatic binding/unbinding gene regulation regime in details. In non-adiabatic regime,  diverse intermediate meta-stable states are emerged in the results of our model. This was only observed previously in the experiments. These results  can help us to understand what controls the epigenetic process in different cell states and also provide physical mechanism for the heterogeneity. This work presents a new way to quantify the cancer heterogeneity from epigenetic perspective. This can provide insights into heterogeneity involved in cancer, SC, and CSC, and may ultimately lead to new approaches to cancer therapy.

\section{Model construction}

We start with a gene regulatory network motif to illustrate the relationship among SC, CS, CSC involving vital regulatory genes and microRNAs. In the motif, there are 6 nodes. P53 and MDM2 are the oncogenes of cancer. OCT4 is a marker gene for stem cell, ZEB is a crucial regulator of EMT during cancer development and two microRNAs (miR145 and miR200) which are vital to the regulations.

\begin{figure*}[!ht]
	\includegraphics[width=16.4cm,height=9cm]{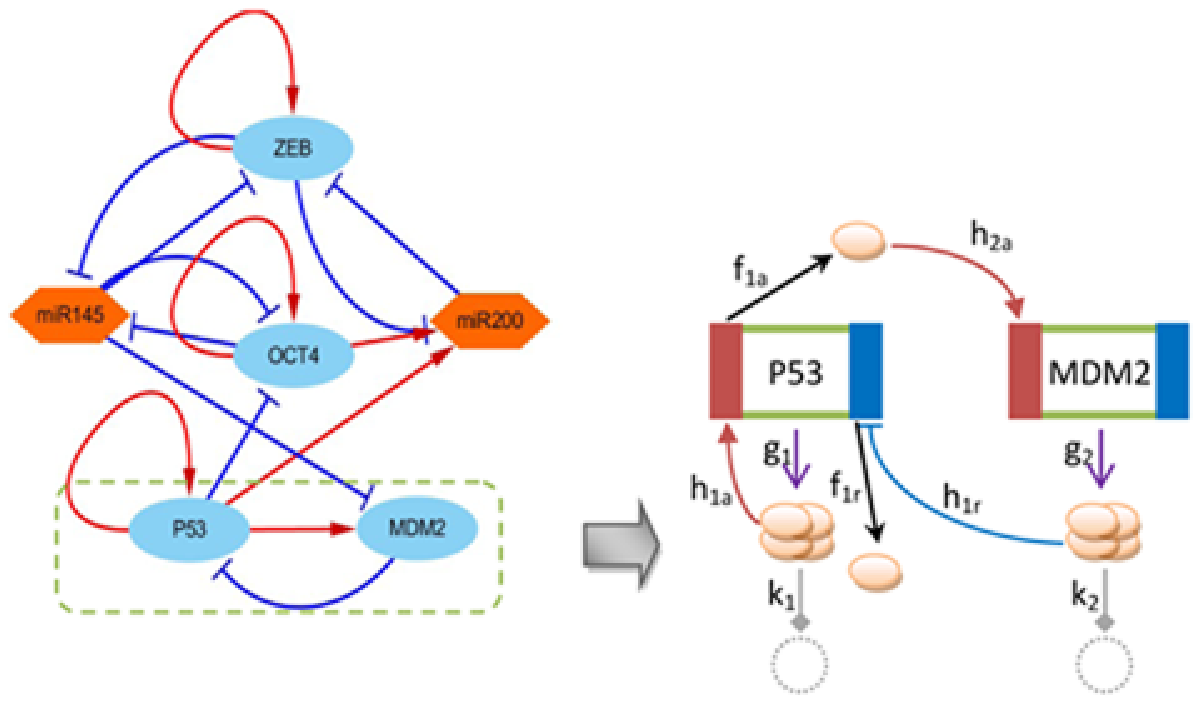}
	\caption{{The regulatory network motif with 6 nodes and 16 regulations.(7 activations and 9 repressions. The arrows represent the activating regulations and the short bars represent the repressing regulations)}\\
		The hexagonal nodes represent the micro-RNAs, the oval nodes represent the genes. The parameter $h$, $f$, $g$ and $k$ represent the protein binding rate, the protein unbinding rate, the protein synthesis rate and the protein degradation rate.} \label{net}
\end{figure*}

In the motif (Fig.\ref{net}) , gene OCT4, P53 and ZEB self activate themselves\cite{Kolesnikoff2014}. OCT4 promotes the transcription of miR-200\cite{Pandey2015} and miR-200 represses the translation of ZEB\cite{Wang2018}. In the meantime, OCT4 inhibits the transcription of miR-145\cite{Liu2017} and miR-145 inhibits the translation of ZEB\cite{Li2018}, OCT4\cite{Liu2017} and MDM2\cite{Teng2018}. The two micro-RNAs (miR-200 and miR-145) inhibit the expression of ZEB\cite{Maleki2018,Kolesnikoff2014} and ZEB also represses the translation of the two micro-RNAs\cite{Li2018}. P53 promotes the expression of MDM2\cite{Freedman1999,AbouJaoud2019} and miR-200 represses the translation of OCT4\cite{Fernandes2019}. MDM2 inhibits the expression of P53\cite{AbouJaoud2019} and is inhibited by miR-145\cite{Teng2018}.

%

The underlying dynamics of the gene regulatory motif is stochastic. One can apply Gillespie algorithm\cite{gillespie1977exact} to quantify the stochastic dynamics and statistical distribution of gene expressions. To describe the biological process precisely, a set of time scale parameters of each process can be defined. The model's chemical reactions can be expressed as follow:
\begin{eqnarray}
\mathcal{G}_A^{0\alpha\beta} + (n+1)B \autorightleftharpoons{$h_{1A}$}{$f_{1A}$} \mathcal{G}_A^{1\alpha\beta} + (n)B
\end{eqnarray}

\begin{eqnarray}
\mathcal{G}_A^{\alpha0\beta} + (n+2)C \autorightleftharpoons{$h_{2A}$}{$f_{2A}$} \mathcal{G}_A^{\alpha1\beta} + (n)C
\end{eqnarray}

\begin{eqnarray}
\mathcal{G}_A^{\alpha\beta0} + (n+4)D \autorightleftharpoons{$h_{3A}$}{$f_{3A}$} \mathcal{G}_A^{\alpha\beta1} + (n)D
\end{eqnarray}

\begin{eqnarray}
(n)\mathcal{G}_A \autorightleftharpoons{$g$}{$k$} (n+1)\mathcal{G}_A
\end{eqnarray}

The parameters $g$ denotes the protein synthesis rate and $k$ denotes the protein degradation rate, $h$ denotes the binding rate and $f$ denotes the unbinding rate of regulatory proteins to the target genes. The protein synthesis rate is influenced by the regulated gene number and regulated type. $\mathcal{G}$ denotes a gene. The gene $A$ has three operator sites. Upper corner mark $\alpha$, $\beta$ of gene $A$ denote a  binding state of a binding site. `0' and `1' denote the unbinding and binding state. Protein B,C and D are monomer, dimer and tetramer respectively. 

In Fig.\ref{net}, we described the regulations of P53 and MDM2. Gene P53 has two operator sites. One is activation binding site, the other is repression binding site. Gene MDM2 has one activation binding site. The gene P53 and MDM2 translated to related proteins with the synthesis rate of $g_1$ and $g_2$. The protein production of P53 binding to the self-activation binding site with the binding rate of $h_{1a}$ which can increase P53 synthesis rate. Protein P53 binding to the promoter of MDM2 with the binding rate of $h_{2a}$. The protein production of MDM2 binding to the repression binding site with the binding rate of $h_{1r}$ which can decease MDM2 synthesis rate. The superscript of $a$ and $r$ represent ``activation" and ``repression" respectively. The protein P53 and MDM2 dissociated from DNA with the unbinding rate of $f_{1a}$ and $f_{1r}$ respectively. The degradation rate of P53 and MDM2 are $k_1$ and $k_2$ respectively . Take P53 as an example, P53 is regulated by two genes in the network motif. One is the self-activation , another is the repression of MDM2. The synthesis rate will be increased or decreased by a factor of $\lambda_a$ or $\lambda_r$, respectively. The synthesis rates of P53 are set as: $g_{00}$, $g_{01}=g_{00}\lambda_a$, $g_{10}=g_{00}\lambda_r$ and $g_{11}=g_{00}\lambda_a\lambda_r$. The index number of $g_{ij}$ denotes the binding site number, and `1' denotes the binding state and `0' denotes the unbinding state. The protein synthesis rate at each time depends on the state of promoters at that time. For simplicity of calculation, we set the unbinding rate as $f$ for all binding sites and the protein degradation rate as $k$ for all protein production. The binding rate depends on the polymeric form of the regulate protein. If the protein is a monomer, the binding rate of the protein A is $h_A= h_An_B$. If the protein is a dimer, the binding rate of the protein A is $h_A= h_An_B(n_B-1)/2$. If the protein is tetramer, the binding rate of the protein A is $h_A= h_An_B(n_B-1)(n_B-2)(n_B-3)/6$. $n_B$ is the protein number of the regulate protein B. We only consider these three cases, because these three polymeric forms are common in the biochemical reactions. 

We define the equilibrium constants: $X_{eq} = f/h$ and the adiabatic parameter: $\omega = f/k$.
$\omega$ is used to describe the relative time scale of the regulatory binding/unbinding to the synthesis/degradation. When the value of $\omega$ is large (termed as adiabatic regime), the regulation processes are relatively fast compared to the synthesis/degradation. This describes the strong coupling regime where proteins once produced are immediately used for the regulations. In other words, the effective gene regulatory interactions are strong. When the value of $\omega$ is small (termed as the non-adiabatic regime), this describes the relatively slow regulation processes compared to the synthesis/degradation, where proteins once produced take some time for the binding regulations. In other words, the effective gene regulatory interactions are weak \cite{Feng2010JPC, Feng2012A, Li2013Quantifying, Chen2016A,Haidong2011Adiabatic}.

\section{Result and Discussion }

\subsection{The heterogeneity from the landscape perspective}

\begin{figure*}[!ht]
	\includegraphics[width=16.4cm,height=12cm]{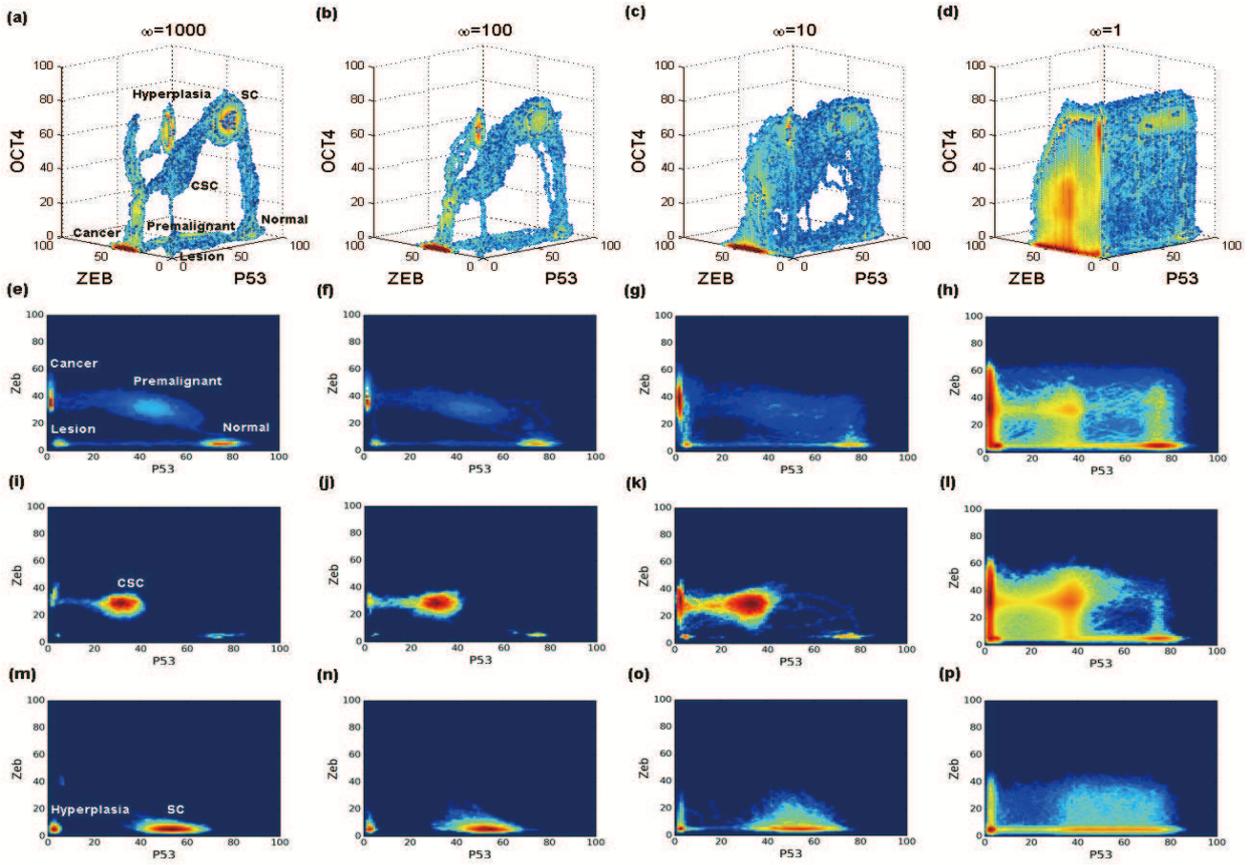}
	\caption{{The comparisons of the 3-dimensional landscape as well as the 2-dimensional slice when the parameter $\omega=1000,100,10,1$ respectively}\\
		(a),(b),(c),(d) show the 3-dimensional landscape when the parameter $\omega=1000,100,10,1$ respectively. (e),(i),(m) show the 2-dimensional slices when $\omega=1000$ and $OCT4=10,40,80$; (f),(j),(n) show the 2-dimensional slices when $\omega=100$ and $OCT4=10,40,80$; (g),(k),(o) show the 2-dimensional slice when $\omega=10$ and $OCT4=10,40,80$ and (h),(l),(p) show the 2-dimensional slice when $\omega=1$ and $OCT4=10,40,80$ respectively} \label{compare}
\end{figure*}

After performing the stochastic simulations on the corresponding gene network with the specified reactions above using Gillespie algorithm \cite{gillespie1977exact}, we can obtain the stochastic trajectories of the genes and associated mRNAs. We can then collect the statistics and obtain the distributions of individual genes and joint distributions among these genes in the long time limit. This provides the quantification of the landscape since probability representing the weight of the state is closely linked to the potential landscape ($U=-lnP$). It is often difficult to visualize the landscape in multi-dimensions, we then choose the gene P53, ZEB and OCT4 to represent cancer, metastasis(EMT) and developmental(stem cell) characteristics. To display the landscape or the weight distribution of the states clearly, we show the comparisons of the 3-dimensional landscapes as well as the 2-dimensional slices. Fig.\ref{compare}(a,b,c,d) show the 3-dimensional landscapes at parameter $\omega=1000,100,10,1$ respectively. Fig.\ref{compare}(e,f,g,h) shows the 2-dimensional slices when $OCT4=10$ at parameter $\omega=1000,100,10,1$ respectively. We can see the variations of Normal, Premalignant, Cancer and Lesion states in the slices. Fig.\ref{compare}(i,j,k,l) shows the 2-dimensional slices when $OCT4=40$ at the parameter $\omega=1000,100,10,1$ respectively. We can see the variations of CSC state in the slices. Fig.\ref{compare}(m,n,o,p) shows the 2-dimensional slice when $OCT4=80$ at the parameter $\omega=1000,100,10,1$ respectively. We can see the variations of Hyperplasia and SC states in the slices.

In Fig.\ref{compare}, we can see that when the adiabatic parameter $\omega$ is set as 1000 (fast binding/unbinding relative to synthesis/degradation) in the strong regulatory coupling adiabatic regime\cite{Feng2011Adiabatic}, seven states are emerged including Normal, Premalignant, Cancer and Lesion, CSC, Hyperplasia and SC states. This corresponds to our previous adiabatic results \cite{Chong2018Quantification}. When the adiabatic parameter $\omega$ becomes smaller, more steady states emerge in this non-adiabatic regime. Because the gene regulation rate is slower than the protein synthesis and degradation, the gene regulations are more weakly coupled.  As a result, multiple metastable-states emerge as the regulatory proteins switch on and off  the target gene less frequently. The slow regulation rate can reflect the non-adiabatic fluctuations and longer time scales of epigenetic effects such as DNA Methylation or histone remodifications\cite{Ashwin2015Effects}. Longer time of binding/unbinding regulations also indicate less chance of coupling among genes at given time intervals. This reduces the effective interactions among genes. As a result of less constraints in terms of the regulatory interactions, more states emerge. The multiple kinetic paths between each two states also emerge and more transitions become available among states.  The emergence of the multiple metastable basin states can provide a physical mechanism for the origin of the heterogeneity \cite{Feng2012A, Li2014JRSI, Chen2016A}.

It is worthwhile to point out that mutations can lead to the heterogeneity by the changes of DNA sequences or the nodes for the gene networks. When mutation is less frequent such as in the cell differentiation and reprogramming process, the heterogeneity can still be significant. This could be due to the epigenetic changes such as histone remodification and DNA methylation which provide extra time scales effectively. This delays the regulation process and therefore weakens the effective regulation strengths. Many states can emerge as a result of the weaker regulations, giving arise to the heterogeneity. The regulation strengths can also be changed directly through the kinetic regulation parameters in the gene networks to explore the cell-cell variability in a population \cite{Kohar2018, Llamosi2016}.

\subsection{The Landscape View of the Cancer and CSC Heterogeneity}

The heterogeneity exists in cancer, CSC and SC which can be observed clearly from this model. In Fig.\ref{compare}, we can see that when the parameter $\omega$ decreases (slower regulation and effectively weaker regulation strength), the heterogeneity of Cancer, CSC and SC states become more and more significant. As the SC heterogeneity is a common phenomenon in mammals\cite{Feng2012A}, we mainly focus on the heterogeneity of cancer \cite{Chen2016A} and CSCs.

The intercellular heterogeneity may result from clonal evolution driven by genetic instability\cite{Shackleton2009Heterogeneity}. This can lead to many different phenotypes and functions. As seen here, the physical mechanism of the intracellular heterogeneity can be due to the weakened regulatory interactions among the genes in the gene network motif \cite{Chen2015Multiple,Kanwal2012Epigenetic}. The epigenetic effects such as DNA methylation, CpG islands promoter hypermethylation, nucleosome remodelling and histone modification can elongate the kinetic process and therefore effectively weaken the gene interactions\cite{Singh2015Evaluation}.

In Fig.\ref{compare}(e), (f), (g) and (h), from left to right, the region of Cancer state basin is enlarged and finally connected to the Lesion state when $\omega$ is decreased to $1$. Experiments showed that accumulations of epigenetic modification such as promoter methylation of the critical genes or DNA repair genes can induce lesions\cite{Hitchins2011Dominantly}. When the regulation rate slows down, the stochastic epigenetic modification in individual cancer cells has more choices of the adaptation and selection to fit the environments. Furthermore, such  evolution may be different in time and space, and different fitness may appear in different environments and stages of the cancer for adaption. Some area may need hypoxia adaption, some area may need fast-growing adaption. When the cancer is developed, the adaption may evolve accordingly, even with the resistance of the drug.

From Fig.\ref{compare}(i), (j), (k) and (l) we can see that, the boundary of CSC state basin gets enlarged which connects to SC and Cancer state hierarchically when $\omega$ is decreased. This hierarchy joins the normal tissues to stem cells and leads to a range of differentiated cancer cells. Importantly, the hierarchical structure which CSC supports is not realized through a one-path route. When the parameter $\omega$ is decreased, the paths became more widely spreaded. Moreover, the paths can be plastic. So the terminally differentiated cancer cells can gain CSCs chracteristics under specific epigenetic conditions. The SCs can gain cancer characteristics and become CSCs. Recently, some studies tracing of CD133+ cells provided direct evidence that SCs were susceptible to cancerous transformation\cite{Zhu2016Multi,Medema2013Cancer}.



\section{Conclusion}
Cancer is a complex and robust disease. The genetic and epigenetic alterations can lead to the cancer heterogeneity. In this study, we studied stochastic processes and associated slow non-adiabatic gene regulatory dynamics. We provide a physical mechanism for heterogeneity from epigenetics. We quantitatively uncovered the heterogeneity of CSC, SC and cancer based on a key gene regulatory network. The heterogeneity has a close relation to the cancer therapy, as the heterogeneity may result various phenotypes in tumor. Understanding the mechanism of heterogeneity in CSC, SC and cancer can help us make  progress in the cancer therapy.

\section{Acknowledgement}

This study was supported by NSFC grant no.91430217, ,MOST-China-Grant No.2016YFA0203200 and grant no.NSF-PHY-76066 and grant NSF-CHE-1808474.

\bibliographystyle{unsrt}

\bf\scriptsize \linespread{0.5} \setlength{\bibsep}{0.0ex}

\bibliography{non}

\clearpage
\clearpage
\clearpage
\end{document}